\documentstyle[12pt,psfig]{article}
\begin{document}

\begin{center}
\bf{NONLOCAL REGULARIZATION FOR NON-ABELIAN}
\end{center}
\begin{center}
\bf{ GAUGE THEORIES }
\end{center}

\begin{center}
\bf{FOR ARBITRARY GAUGE PARAMETER}
\end{center}

\begin{center}
Anirban Basu
\end{center}

\begin{center}
and
\end{center}

\begin{center}
Satish D. Joglekar
\end{center}

\begin{center}
Department of Physics
\end{center}

\begin{center}
I.I.T. Kanpur
\end{center}

\begin{center}
Kanpur 208016, INDIA
\end{center}

{\bf{ABSTRACT}}\\
~~~~~~~We study the nonlocal regularization for the non-abelian gauge theories
for an arbitrary value of the gauge parameter \( \xi  \). We show that
the procedure for the nonlocalization of field theories established earlier
by the original authors, when applied in that form to the Faddeev-Popov effective
action in a linear gauge cannot lead to a \( \xi  \)-independent result for the
observables. We then show that an alternate procedure which is simpler can be
used and that it leads to the S-matrix elements (where they exist) independent
of \( \xi  \).

~~~~~~~ ~~~~~~~~~~~

{\bf{1.INTRODUCTION}}\\
 ~~~Local Quantum Field Theories are plagued with infinities and need regularization
to make the process of renormalization mathematically well-defined. Many regularizations
have been proposed over the last 50 years, dimensional regularization being
one used widest due to its effectiveness{[}1{]}. While dimensional regularization
is useful in a wide class of Quantum Field Theories, it cannot be used directly
in Supersymmetric Field Theories. A number of regularizations have been proposed
over the last decade that can be used in Supersymmetric Field Theories{[}2,3{]}.
Nonlocal regularization is one of them{[}2, 4, 5{]}.

Nonlocal regularization proposed by Evans et al{[}2{]} has been extensively
studied{[}\( 4,5,6 \){]}. Renormalization procedure has been established
upto two loop order{[}5{]} in scalar theories. The scheme has found
an elegant and neat formulation in reference  4 which has shown how
nonlocally regularized field theories can be constructed from a local QFT in
a systematic fashion. More importantly, it has been established that local/global
symmetries can be preserved in their nonlocal form and the WT identities of
local QFT's derivable from local symmetries such as gauge invariance/BRS symmetry
find their natural nonlocal extensions. This has been done for the abelian gauge
theories to all orders{[}7{]} and for nonabelian gauge theories in Feynman
gauge{[}4{]} upto one loop order {[}limited only by the existence
of measure beyond one loop{]}.

Nonlocally regularized theories have also found other equally useful\\
interpretations{[}4, 8{]}.
Nonlocally regularized theories contain in them a large mass parameter \( \Lambda  \).
It has been shown (wherever the measure factor exists) that these theories are
unitary even with a finite \( \Lambda  \). Discussions of causality and renormalization
group have also been carried out{[}4, 5{]}. Thus it has been suggested{[}8,9,10{]}
such nonlocally regularized theories with a finite \( \Lambda  \) can themselves
be looked upon as valid physical theories (rather than a regularization for
which \( \Lambda \rightarrow \infty  \) must be taken). The parameter \( \Lambda  \)
has been interpreted in two ways: (a) as a signal of an underlying space-time
granularity; (b) as the mass scale beyond which the physical theory must be
replaced by another, more fundamental theory. We may regard view (a) as a mathematically
convenient way of embodying space-time granularity in QFT's in a way that is
physically consistent. In view (b), we may regard the nonlocal QFT as an effective
field theory that may have been derived from a more fundamental theory beyond
the scale \( \Lambda  \). Thus, for example, we regard \textbf{nonlocal} standard
model as the effective theory of fundamental processes at present energies,
in which a signature of physics beyond standard model and the scale at which
the SM should break down are both implicit in the scale \( \Lambda  \). An
attempt to put lower bound using (g-2) of the muon has been made in Ref. 8.

The setting of such nonlocal QFT's has also been used to understand renormalization
program in a mathematically rigorous way{[}10{]}. A way to put an upper bound
on \( \Lambda  \) has also been suggested{[}9, 10{]}.

Nonlocal regularization has also found use in the discussion of higher loop
anomalies in BV formulation{[}11{]}.

In view of the above, it seems valuable to study these formulations further.
One of the features of linear gauges in local gauge theories is the availability
of a free parameter \( \xi  \) (gauge parameter) which helps in verifying the
gauge independence of physical results. \( \xi  \)-independence of physical
results in spontaneously broken gauge theories has also been used to establish the cancellation of contributions 
from the unphysical poles to the cutting equations in SBGT[12].\\

It is therefore desirable that we have a formulation of non-local nonabelian
gauge theories valid for an arbitrary \( \xi  \). Now, a well laid-out procedure
for the nonlocalization of field theories has been presented in references 4
and 5. We found however that when we applied this procedure to the spontaneously
broken theory (SM) in \( R_{\xi } \) gauges and calculated the (g-2) for the
muon{[}8{]} we found a \( \xi  \)-dependent result {[}13{]}. This motivated
us to look into the question of nonlocal formulation of unbroken nonabelian
gauge theories and of spontaneously broken chiral abelian gauge theories{[}14{]}.
In the present work, we concern ourselves with the former.\\ We now discuss the
plan of our work.
In Section II, we summarize the results on the nonlocal quantum field theories
of 2, 4, 5. In Section III, we adopt the procedure outlined in
4 for nonlocalization for arbitrary \( \xi  \) and evaluate \( \xi \frac{dW}{d\xi }|_{\xi =1} \)for
this case. We obtain a term in \( \xi \frac{dW}{d\xi }|_{\xi =1} \) that $\bf{can}$
contribute to on-shell physical processes and which cannot be cancelled by a
\( \xi  \) dependent measure. In Section IV, we suggest an alternate way of
constructing nonlocal unbroken gauge theories for an arbitrary \( \xi  \) and
establish the WT identity satisfied by the physical Green's function \( \xi \frac{dW}{d\xi }|_{J=J_{phy}} \)
of eqn. (4.15). This equation is analogous to that in that in the local case;
and should lead to the \( \xi  \)-independence of physical quantities that
are free of infra-red divergences.

\textbf{II REVIEW OF KNOWN RESULTS}\\
\\
\textit{\bf{A~Nonlocal Regularization}}\\
\\
 Let us briefly review the method of non-local regularization as proposed in
{[}4, 5{]}. Let \( \phi _{i} \) stand for a generic, not necessarily
scalar, field and let us assume that the local action can be written as a standard
free part plus an interaction\\
\\
 S{[}\( \phi  \){]} = F{[}\( \phi  \){]} + I{[}\( \phi  \){]} ,\hspace{2in}(2.1)\\
\\
 where\\
\\
 F{[}\( \phi  \){]} = \( \frac{1}{2} \) \( \int  \) \( d^{D}x \) ~\( \phi _{i} \)\( (x) \)~\( {\mathcal{F}}_{ij} \)
~\( \phi _{j} \)\( (x) \)~~~~~~~~~~~~~~~~~~~~~~~~~~~~~~~~~~(2.2)\\
 \\
Here, ${\mathcal{F}}$ is the kinetic energy operator for the field \( \phi  \), and I{[}\( \phi  \){]}
is the interaction term. For unbroken gauge theories, S{[}\( \phi  \){]} would be the
BRS gauge fixed action and \( \phi _{i} \) would include both the fields of
the invariant action and the ghosts introduced in the process of fixing the
gauge. \\
\\
 From the kinetic energy operator \( {\mathcal{F}} \), let us define a non-local smearing
operator \( {\mathcal{E}} \) and a shadow kinetic operator \( {\mathcal{O}}^{-1} \) as\footnote{In the case of the ghost Lagrangian, its overall sign (and therefore that of the quadratic form) is arbitrary. In such a case, we shall always take the sign of ${\mathcal{F}}$ such that the ghost propagator is damped for large $\mid$$k^2$$\mid$ in Euclidean space. }\\
\\
 \( {\mathcal{E}} \) = exp{[}\( \frac{{\mathcal{F}}}{2\Lambda ^{2}} \){]}~~~~~~~
~~~~~~~~~~~~   (2.3)\\
 and \\
\\
 \( {\mathcal{O}} \) =~\( \frac{{\mathcal{E}}^{2}-\, 1}{\mathcal{F}} \)~~~~~~~
~~(2.4)\\
\\
 Further, the smeared field is defined as \\
\\
 \( {\hat{\phi}}  \) = \( {\mathcal{E}}^{-1} \) \( \phi  \)~~~~~~~~~~~~~~
~(2.5)\\
\\
For every field \( \phi  \), an auxiliary field \( \psi  \) of the same type
is introduced. Then, the auxiliary action is defined to be \\
\\
\( {\mathcal{S}} \){[}\( \phi , \)\( \psi  \){]} = F{[}\( {\hat{\phi}}  \){]} - A{[}\( \psi  \){]}
+ I{[}\( \phi + \)\( \psi  \){]}~~~~~~~~~~~~~~~~~~~~~~~(2.6)\\
\\
where\\
\\
A{[}\( \psi  \){]} = \( \frac{1}{2} \)\( \int  \) \( d^{D}x \) ~\( \psi _{i} \)\( (x) \)~\( {\mathcal{O}}^{-1}_{ij} \)
~\( \psi _{j} \)\( (x) \)~~~~~~~~~~~~~~~~~~~~(2.7)\\
\\
The action for the nonlocalized theory ${\hat{S}}${[}\( \phi  \){]} is defined to be
\\
\\
${\hat{S}}${[}\( \phi  \){]} =$ {\mathcal{S}}${[}\( \phi , \)\( \psi  \){[}\( \phi  \){]}{]}~~~~~~~~~~~~~~~~~~~~
(2.8)\\
\\
where \( \psi  \){[}\( \phi  \){]} is a solution of the classical shadow field
equation\\
\\
\( \frac{\delta {\mathcal{S}}[\phi ,\psi ]}{\delta \psi _{i}} \) = 0~~~~~~~(2.9)\\
\\
Quantization is carried out in the path integral formulation. The quantization
rule is\\
\\
\( <T^{*}(O[\phi ])>_{\mathcal{E}} \) = \( \int [D\phi ] \)\( \mu  \){[}\( \phi  \){]}\( O[{\hat{\phi}} ] \)\( e^{i{\hat{S}}[\phi ]} \)~~~~~~~~~~~(2.10)\\
\\
Here O is any operator taken as a functional of fields. \( \mu  \){[}\( \phi  \){]}
is the measure factor defined such that \( [D\phi ] \)\( \mu  \){[}\( \phi  \){]}
is invariant under the nonlocal generalisation of the local symmetry. For nonlocalized
non-Abelian gauge theories, this measure factor can be non-trivial and has been
evaluated upto one loop {[}4{]}. For the abelian gauge theories, the measure factor is known to all orders[7].\\
\\
The nonlocalized Feynman rules are simple extensions of the local ones. The
vertices are unchanged but every leg can connect either to a smeared propagator
\\
\\
\( \frac{i{\mathcal{E}}^{2}}{{\mathcal{F}}+i\epsilon } \) = -i\( \int _{1}^{\infty }\frac{d\tau }{\Lambda ^{2}} \)\( e^{\frac{{\mathcal{F}}\tau }{\Lambda ^{2}}} \)~~~~~~~~~~~~~~~~~(2.11)\\
\\
or to a shadow propagator\\
\\
\( \frac{i(1-{\mathcal{E}}^{2})}{{\mathcal{F}}} \) = -i${\mathcal{O}}$ = -i\( \int ^{1}_{0}\frac{d\tau }{\Lambda ^{2}} \)\( e^{\frac{{\mathcal{F}}\tau }{\Lambda ^{2}}} \)~~~~~~~~~~~~~~~~~~~
(2.12)\\
\\
Diagramatically, they will be represented as \\
 \\
 the smeared or ``unbarred'' propagator\\
\\
\begin{center}
\raisebox{0.0cm}{\psfig{file=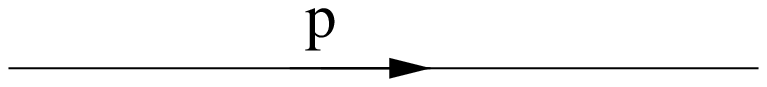}}
\end{center}

the shadow or ``barred" propagator\\

\begin{center}
\raisebox{0.0cm}{\psfig{file=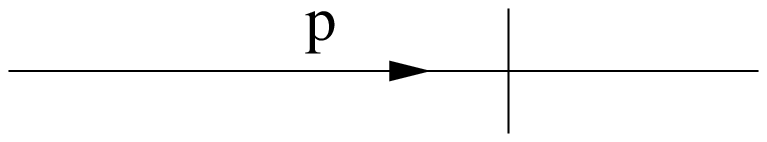}}
\end{center}

The shadow propagator lacks a pole and so carries no quanta. Thus, the following
points are to be noted:\\
\\
i) All external lines must be unbarred.\\
ii) The symmetry factor for any diagram is computed without distinguishing between
barred and unbarred lines. \\
iii) The loop integrations are well defined in the Euclidean space because of
the exponential damping factors coming from propagators within loops.\\
iv) Internal lines can be smeared or barred. However, loops containing only the shadow lines are forbidden.
\\
v) Tree order Green's functions are unchanged except for external line factors
which are unity on shell. This follows because every internal line of a tree
graph can be either barred or unbarred. Hence, it is the sum of both these that
which enters, which gives the local propagator.\\
\\
{\bf{B~ Theorems Regarding Nonlocal Regularized Actions}}\\
\\
Before discussing the nonlocal BRS symmetries, let us consider a few theorems
concerning classical solutions of the Euler-Lagrange equations associated with
the local action S{[}\( \phi  \){]}, the auxiliary action ${\mathcal{S}}${[}\( \phi ,\psi  \){]}
and the nonlocalized ${\hat{S}}${[}\( \phi  \){]} action:\\
\\
\textit{Theorem A.1}: The shadow fields can be expressed as follows:\\
\\
\( \psi _{i}[\phi ] \) = -\( (\frac{{\mathcal{E}}^{2}-1}{{\mathcal{E}}^{2}})_{ij} \)\( \phi _{j} \)
+ \( O_{ij} \)\( \frac{\delta {\hat{S}}[\phi ]}{\delta \phi _{j}} \)~~~~~~~~~~~~
~~(2.13)\\
\textit{}\\
\textit{Theorem A.2:} If \( \phi _{i} \) and \( \psi _{i} \) obey the Euler-Lagrange
equations of ${\mathcal{S}}${[}\( \phi ,\psi  \){]} then \( \chi _{i}= \) \( \phi _{i} \)
+ \( \psi _{i} \) obeys the Euler-Lagrange equations of S{[}\( \chi  \){]}.\\
\\
\textit{Theorem A.3}: If \( \chi _{i} \) obeys the Euler-Lagrange equations
of S{[}\( \chi  \){]}, then the following fields\\
\\
\( \phi _{i} \) = \( {\mathcal{E}}^{2}_{ij} \)\( \chi _{j} \)~~~~~~~~~~~~~~~~   (2.14)\\
\( \psi _{i}= \) \( (1-{\mathcal{E}}^{2})_{ij} \)\( \chi _{j} \)\\
\\
obey the Euler-Lagrange equations of ${\mathcal{S}}${[}\( \phi ,\psi  \){]}.\\
\\
Let us also consider another set of theorems concerning classical symmetries
of S{[}\( \chi  \){]}, ${\mathcal{S}}${[}\( \phi ,\psi  \){]} and ${\hat{S}}${[}\( \phi  \){]}:\\
\\
\textit{Theorem B.1}: If S{[}\( \phi  \){]} is invariant under the infinisitesimal
transformation \\
\\
\( \delta \phi _{i}= \) \( T_{i}[\phi ] \), \\
\\
then the following transformation is a symmetry of ${\mathcal{S}}${[}\( \phi ,\psi  \){]}:\\
\\
\( \Delta \phi _{i}= \) \( {\mathcal{E}}^{2}_{ij} \)\( T_{j}[\phi +\psi ] \) ~~~~~~~~~~~~~~~~~~                        (2.15)\\
\( \Delta \psi _{i} \) = \( (1-{\mathcal{E}}^{2})_{ij}T_{j}[\phi +\psi ]. \)\\
\\
\textit{Theorem B.3}\footnote{We stick to the theorem numbering of [4].}: If S{[}\( \phi  \){]} is invariant under \( \delta \phi _{i}= \)
\( T_{i}[\phi ] \), then ${\hat{S}}${[}\( \phi  \){]} is invariant under\\
\\
\({\hat{ \delta}} \phi _{i} \) = \( {\mathcal{E}}^{2}_{ij}T_{j}[\phi +\psi [\phi]] \)~~~~~~~~~~~~~~~~~~~~~~(2.16)\\
\\
\textit{TheoremB.4}: The following transformation generates a symmetry of ${\mathcal{S}}${[}\( \phi ,\psi  \){]}:\\
\\
\( \Delta \phi _{i} \) = \( A_{ij}[\phi ,\psi ]\{\frac{\delta {\mathcal{S}}[\phi ,\psi ]}{\delta \phi _{j}}-\frac{\delta {\mathcal{S}}[\phi ,\psi ]}{\delta \psi _{j}}\} \) ~~~~~~~~~~~~~~~~~~~~~~~~~~~~                        (2.17)\\
\( \Delta \psi _{i} \) = -\( A_{ij}[\phi ,\psi ]\{\frac{\delta {\mathcal{S}}[\phi ,\psi ]}{\delta \phi _{j}}-\frac{\delta {\mathcal{S}}[\phi ,\psi ]}{\delta \psi _{j}}\} \)\\
\\
provided \( A_{ij}[\phi ,\psi ] \) = -\( A_{ji}[\phi ,\psi ] \). This symmetry is a "trivial symmetry'' without a dynamical content. \\
\\
\( \Delta \phi _{i} \) can also be cast in the simple form \\
\\
\( \Delta \phi _{i} \) = \( A_{ij}[\phi ,\psi ] \)\( {\mathcal{E}}^{-2}_{jk}{\mathcal{O}}^{-1}_{kl}[(1-{\mathcal{E}}^{2})_{lm}\phi _{m}-{\mathcal{E}}^{2}_{lm}\psi _{m}] \)~~~~~~~~~~~~~~~~~~~~(2.18)\\
\\
An important special case is given by the choice\\
\\
\( A_{il}[\phi ,\psi ] \) = \( M_{ij}{\mathcal{O}}_{jk}{\mathcal{E}}^{2}_{kl} \)\\
\\
where\\
\\
\( [M_{ij},\, {\mathcal{F}}_{ij}] \) = 0 and \( M_{ij}= \) \( -M_{ji} \).\\
\\
We next review the nonlocal BRS symmetries of the Yang-Mills theory using the above results.\\
\\
{\bf{C Nonlocal Regularization of Yang-Mills Theory in the Feynman
Gauge}}\\
\\
Finally, let us consider the nonlocal regularization of Yang-Mills theories.
First, we will study the results obtained in the Feynman gauge. \\
\\
The Feynman gauge local BRS Lagrangian is \\
\\
\( {\mathcal{L}}_{BRS} \)=-\( \frac{1}{2} \)\( \partial _{\mu }A_{\nu }^{a}\, \partial ^{\mu }A^{a\nu } \)-\( \partial _{\mu } \)\( {\overline{\eta}} ^{a}\, \partial ^{\mu }\eta ^{a} \)+\( g\, f^{abc}\partial _{\mu }{\overline{\eta}} ^{a}\, A^{b\mu }\eta ^{c} \)+\( g\, f^{abc}\partial _{\mu }A_{\nu }^{a}\, A^{b\mu }A_{c\nu } \)\\
-\( \frac{g^{2}}{4} \)\( f^{abc}f^{cde}A^{a}_{\mu }A^{b}_{\nu }A^{d\mu }A^{e\nu } \)~~~~~~~~~~~~~~~~~~~~~~(2.19)
\\
Thus, the gluon and the ghost kinetic energy operators are \( {\mathcal{F}}^{\mu \nu }_{ab}= \)
\( \delta _{ab}\eta ^{\mu \nu }\partial ^{2} \) and \( {\mathcal{F}}_{ab} \) = \( -\delta _{ab}\partial ^{2} \)
respectively\footnote{See earlier footnote above eq.(2.3).}. \\
\\
Let us denote the auxiliary fields of ${A_{\mu}^a}$ and ${{\eta}^a}$ by ${B_{\mu}^a}$ and ${{\psi}^a}$ respectively.\\
\\
Thus the non-localized BRS action is\\
\\
${\hat{S}}${[}A,\( \eta ,{\overline{\eta}}  \){]}=\( \int  \)\( d^{4}x \)\{-\( \frac{1}{2} \)\( \partial _{\mu }{\hat{A}}^{a}_{\nu }\, \partial ^{\mu }{\hat{A}}^{a\nu } \)-\( \frac{1}{2} \)\( B^{a}_{\mu }O^{-1}B^{a\mu } \)-\( \partial _{\mu }{{\hat{\overline{\eta}}}} ^{a}\, \partial ^{\mu }{\hat{\eta}} ^{a} \)+\( \psi ^{a}O^{-1}\psi ^{a} \)\}\\
+I{[}A+B,\( \eta +\psi  \),\( {\overline{\eta}} +{\overline{\psi}}  \){]}~~~~~~~~~~~~~~~~~~~(2.20)\\
\\
The local BRS Yang-Mills action in the Feynman gauge has the following BRS symmetry
transformations:\\
\\
 \( \delta A_{\mu }^{a} \) = \( (\partial _{\mu }\eta ^{a}-gf^{abc}A^{b}_{\mu }\eta ^{c})\delta \varsigma  \)\\
\\
\( \delta \eta ^{a} \) = -\( \frac{g}{2} \)\( f^{abc}\eta ^{b}\eta ^{c}\delta \varsigma  \)~~~~~~~~~~~~~~~~~(2.21)\\
\\
\( \delta {\overline{\eta}} ^{a}=-\partial _{\mu }A^{a\mu }\delta \varsigma  \)\\
\\
where \-\( \delta \varsigma  \) is a constant anticommuting C number.\\
\\
Given the local symmetry transformations for the fields, one can easily write
down their non-local counterparts:\\
\\
\( {\hat{\delta}} A_{\mu }^{a} \) = \( {\mathcal{E}}^{2} \)\( [\partial _{\mu }(\eta ^{a}+\psi ^{a})-gf^{abc}(A^{b}_{\mu }+B^{b}_{\mu })(\eta ^{c}+\psi ^{c})]\delta \varsigma  \)\\
\\
\( {\hat{\delta}} \eta ^{a} \) = -\( \frac{g}{2} \)\( f^{abc}{\mathcal{E}}^{2}(\eta ^{b}+\psi ^{b})(\eta ^{c}+\psi ^{c})\delta \varsigma  \)~~~~~~~~~~~~~(2.22)\\
\\
\( {\hat{\delta}} {{\overline{\eta}} ^{a}}=-{\mathcal{E}}^{2}\partial _{\mu }(A^{a\mu }+B^{a\mu })\delta \varsigma  \)\\
\\
where\\
\\
\({\mathcal{E}} \) = \( e^{\frac{\partial ^{2}}{2\Lambda ^{2}}} \)\\
\\
In [4], it was found that it is convenient to construct a modified nonlocal BRS symmetry transformation by adding a "trivial''
symmetry transformation to the kind (2.18). This was so as noted in [4] since it yielded a variation of ${\overline{c}}$ proportional to $\partial$.A. Put alternatively, we find that these new transformations of [4] have two useful properties 
(i) $\partial$.$\delta$A is directly reducible in terms of the ghost action, (ii) WT identities so formulated allow an easy evaluation of $\xi$$\frac{\partial W}{\partial \xi}$. The measure factor is defined with respect to the latter transformations. They are\\
\\
\( {\hat{\delta}} A_{\mu }^{a} \) = \( [\partial _{\mu }\eta ^{a}-gf^{abc}{\mathcal{E}}^{2}(A^{b}_{\mu }+B^{b}_{\mu })(\eta ^{c}+\psi ^{c})]\delta \varsigma  \)\\
\\
\( {\hat{\delta}} \eta ^{a} \) = -\( \frac{g}{2} \)\( f^{abc}{\mathcal{E}}^{2}(\eta ^{b}+\psi ^{b})(\eta ^{c}+\psi ^{c})\delta \varsigma  \)~~~~~~~~~~~~~(2.23)\\
\\
\( {\hat{\delta}} {{\overline{\eta}} ^{a}}=-\partial _{\mu }A^{a\mu }\delta \varsigma  \)\\
\\
The measure factor[4] is\\
\\
ln($\mu[A,\eta,\overline{\eta}]$)=-$\frac{g^2}{2} f_{acd} f_{bcd} \int {d^{D}}x A_{\mu a} \mathcal{M} $${A_{b}}^{\mu}$ + O($g^{3}$)~~~~(2.24)\\
\\
where\\
\\
$\mathcal{M}$=$\frac{1}{2^{D} {\pi}^{\frac{D}{2}}} \int_{0}^{1}d \tau \frac{\Lambda^{D-2}}{{(\tau + 1)}^{\frac{D}{2}}}$  exp$(\frac{\tau}{\tau + 1} \frac{\partial^{2}}{\Lambda^{2}}) [\frac{2}{\tau + 1} -(D-1) +2(D-2)\frac{\tau}{\tau +1}]$\\
\\
{\bf{III Difficulty with the method of non-local regularization for an arbitrary  $\xi$:}}\\
\\
The above method of regularization worrks correctly in the Feynman gauge \( \xi  \)=1.
The regulator operators are simple and calculations have been performed in this
gauge with relative ease {[}4, 8{]}. When this procedure of taking the
entire quadratic form ${\mathcal{F}}$ which enters the regulator operators is used, the regulators
are \( \xi  \)-dependent and complicated. This, of course, is not a serious
objection to the use of this procedure in {[}4, 5{]}, we find
that the procedure in fact leads to WT identities which imply that the S-matrix
elements, where they exist, are not \( \xi  \)-independent even in one loop
order. In this section, we wish to demonstrate it and then suggest, in the next
section, an alternate way of regularization which is at once simpler and leads
to a WT identity which formally implies the \( \xi  \)-independence of the S-matrix.\\

An abelian special case of this has already been applied to QED[7].\\
Originally we derived the motivation for this work from the following observation
in the context of the SM (where physical S-matrix elements generally exist).
We had found \( \xi  \)-dependence of the muon anomalous magnetic moment in
the SM in {[}8{]} when we follow the procedure of references[4, 5]  as applied to
the spontaneously broken (local) theory for the \( R_{\xi } \) gauges[13]. The
discussion given here in this work has also been extended {[}14{]} to the spontaneously
broken U(1) chiral model where a similar \( \xi  \)-dependence of a physical
quantity has neen demonstrated. The procedure, formulated here in section IV
has been applied there to this case; with formal \( \xi  \)-independence established{[}14{]}.
\\
\\
\\
We now consider the non-local action for an arbitrary \( \xi  \). Here, we
will generalize (following {[}4, 5{]} ), for an arbitrary \( \xi  \),
the appropriate non-local action. We express\\
\\
\( {\hat{S}}_{\xi } \)=\( {\hat{S}}+\Delta {\hat{S}} \), ~~~~~~~~~~~~~~~~~~~(3.1)\\
\\
where\\
\\
${\hat{S}}$=\( \int d^{4}x(-\frac{1}{2}\partial _{\mu }{\hat{A}}_{\nu }^{a}\, \partial ^{\mu }{\hat{A}}^{a\nu }-\partial _{\mu }{\hat{\overline{\eta}}} ^{a}\, \partial ^{\mu }{\hat{\eta}} ^{a}-\frac{1}{2}B_{\mu }^{a}{\mathcal{O}}_{A}^{-1ab\mu \nu }B_{\nu }^{b}+{\overline{\psi}} ^{a}{\mathcal{O}}_{\eta }^{-1ab}\psi ^{b} \)
\\
+Interaction terms)~~~~~~~~~~~~(3.2)\\
\\
and \\
\\
\( \Delta {\hat{S}} \)=\( \frac{1}{2}(1-\frac{1}{\xi })\int d^{4}x(\partial .{\hat{A}}^{a})^{2} \)~~~~~~~~~~~~~~~~~~~~~~~~(3.3)\\
\\
We note that as \( \Lambda \rightarrow \infty  \), ${\hat{S}}$ reduces to the local action
of the Feynman gauge. Note, now, that the smeared gauge field ${\hat{A}}$ has been constructed
using the full (\( \xi  \) dependent) quadratic form ${\mathcal{F}}$\\
\\
\( {\hat{A}}_{\mu }={\mathcal{E}}^{-1}_{A\mu \nu }A^{\nu }=(e^{-\frac{{\mathcal{F}}}{2\Lambda ^{2}}})_{\mu \nu }A^{\nu } \)~~~~~~~~~~~~~~~~~(3.4)\\
\\
and the ghost fields \( {\hat{\eta}}  \) and \( {\hat{\overline{\eta}}}  \) have been smeared using their
respective quadratic forms\\
\\
\( {\hat{\eta}} ={\mathcal{E}}_{\eta }^{-1}\eta  \)~~~~~~~~\( {\hat{\overline{\eta}}} ={\mathcal{E}}_{\eta }^{-1} \)\( {\overline{\eta}}  \)~~~~~~~~~~~~~~~~\( {\mathcal{E}}_{\eta }=e^{\frac{-\partial ^{2}}{2\Lambda ^{2}}} \)
~~~~~~~~~~~~~~~(3.5)\\
\\
where it should be noted that ${\mathcal{E}}_{\eta}$ is independent of \( \xi  \).\\
\\
We shall now proceed to evaluate \( \xi \frac{\partial W}{\partial \xi }|_{\xi =1} \)
for the above nonlocal theory. We note that the ~\( \xi  \) dependence of W
comes from (i) the explicit ~\( \xi  \) dependence of \( \Delta {\hat{S}} \) (ii)
the implicit~\( \xi  \) dependence of \( {\mathcal{E}}_{A} \) in ${\hat{A}}$ (iii) the explicit
\( \xi  \) dependence of \( {\mathcal{O}}^{-1} \) in the B-field kinetic energy term
and (iv) the implicit \( \xi  \) dependence of auxiliary fields B, \( \psi ,\, {\overline{\psi}}  \) and (v) finally from the measure $\mu$($\xi$).
Of these, contribution (iv) vanishes since the auxiliary fields satisfy
\( \frac{\delta S}{\delta \psi }|_{\psi =\psi [\phi ]} \)=0.\\
\\
The first contribution reads\\
\\
(I)=$\langle$$\langle$\( \frac{i}{2\xi }\int d^{4}x(\partial .{\hat{A}}^{a})^{2} \)$\rangle$$\rangle$=$\langle$$\langle$\( \frac{i}{2\xi }\int d^{4}x\, \partial .A^{a}\, {\mathcal{E}}_{R}^{-2}\, \partial .A^{a} \)$\rangle$$\rangle$~~~~~~~~~~~~~(3.6)\\
\\
{[}with \( {\mathcal{E}}_{R}^{-2}=e^{\frac{\partial ^{2}}{\Lambda \xi }} \){]}\\
\\
The implicit ~~\( \xi  \) dependence of ${\hat{A}}$ in \( \Delta S \) contributes \\
\\
(IIA)=-\( \frac{i(1-\xi )}{2}<<\int d^{4}x\frac{\partial }{\partial \xi }(\partial .{\hat{A}}^{a})^{2}>> \)~~~~~(3.7)\\
~~~~~=-\( \frac{i(1-\xi )}{2\Lambda ^{2}\xi ^{2}}<<\int d^{4}x\, (\partial .{\hat{A}}^{a})\, e^{-\frac{\partial ^{2}}{2\Lambda ^{2}\xi }}\, \partial ^{2}(\partial .A^{a})>> \)~~~~~~~~~~~~~(3.8)\\
\\
and it vanishes at~ ~\( \xi  \)=1.\\
\\
The contribution from the implicit dependence on~\( \xi  \) of ${\hat{A}}$ and \( {\mathcal{O}}^{-1} \)
in the B-field kinetic terms can be computed straightforwardly. The result reads\\
\\
(III)=-\( \frac{i}{2\Lambda ^{2}\xi }<<\int d^{4}x\, (\partial .{\hat{A}}^{a})\, e^{-\frac{\partial ^{2}}{2\Lambda ^{2}\xi }}\, \partial ^{2}(\partial .A^{a})>> \)+\\
\( \frac{i}{2\xi }<<\int d^{4}x\, (\partial .B^{a})\frac{1}{1-e^{-\frac{\partial ^{2}}{\Lambda ^{2}\xi }}}(\partial .B^{a})>> \)-\\
\( \frac{i}{2\Lambda ^{2}\xi ^{2}}<<\int d^{4}x\, (\partial .B^{a})\, e^{-\frac{\partial ^{2}}{\Lambda ^{2}\xi }}\, \partial ^{2}(\frac{1}{1-e^{-\frac{\partial ^{2}}{\Lambda ^{2}\xi }}})^{2}\, (\partial .B^{a})>> \)~~~~~~~~~~~~~~~(3.9)\\
\\
The \( (\partial .{A^a})^{2} \) type terms in I and III combine to give\\
\\
\( \frac{i}{2\xi }\int d^{4}x\, (\partial .A^{a})\, {\mathcal{E}}_{R}^{-2}[1-\frac{\partial ^{2}}{\Lambda ^{2}\xi }]\, (\partial .A^{a}) \)~~~~~~~~~~~~~~~~(3.10)\\
\\
At \( \xi  \)=1, these can be simplified using the identity (A.5) in Appendix derived
using the BRS WT identity. We note that as far as Green's functions with external
gauge fields are concerned, we can set terms \( \sim  \)$\langle$$\langle$\( P(\xi ){\overline{\eta}} ^{a}\frac{\delta S}{\delta {\overline{\eta}} ^{a}} \)$\rangle$$\rangle$
to zero as shown in III of the appendix.We are then left with\\
\\
 \( -\frac{i}{2}\int d^{4}x\, {\mathcal{E}}_{R}^{-2}[1-\frac{\partial ^{2}}{\Lambda ^{2}\xi }]\, \partial .A^{a}(x) \)\( {\overline{\eta}} ^{a}(x) \)~\( \int d^{4}y\, J^{b\mu }(\partial _{\mu }\eta ^{b}-gf^{bcd}{\mathcal{E}}^{2}(A^{c}_{\mu }+B^{c}_{\mu })(\eta ^{d}+\psi ^{d})) \)~~~~~~(3.11)\\
\\
Next we simplify the \( (\partial .B^{a})^{2} \) type terms using the relation\\
(\( \partial .B^{a} \))=\( \frac{e^{\frac{\partial ^{2}}{\Lambda ^{2}\xi }}}{\Lambda ^{2}}[\partial .\frac{\delta {\hat{S}}_{\xi }}{\delta A^{a}}-\frac{e^{\frac{\partial ^{2}}{\Lambda ^{2}\xi }}}{\xi }\partial ^{2}(\partial .A^{a})] \)~~~~~~~(3.12)\\
\\
We note that these terms together simplify to yield\\
\\
\( \xi \frac{\partial W}{\partial \xi }=\frac{i}{2\xi }\int d^{4}x<<\partial .\frac{\delta {\hat{S}}_{\xi }}{\delta A^{a}}{\hat{P}}_{1}(\xi )\partial .\frac{\delta {\hat{S}}_{\xi }}{\delta A^{a}}>> \)\\
      -\( \frac{i}{\xi }\int d^{4}x<<\partial .\frac{\delta {\hat{S}}_{\xi }}{\delta A^{a}}{\hat{P}}_{2}(\xi )\partial .A^{a}>> \)\\
          +\( \frac{i}{2\xi }\int d^{4}x<<\partial .A^{a}{\hat{P}}_{3}(\xi )\partial .A^{a}>> \)~~~~~~~~~~~~~(3.13)\\
\\
 where\\
 \\
\( {\hat{P}}_{1}(\xi )={\hat{P}}(\xi )\frac{e^{\frac{2\partial ^{2}}{\Lambda ^{2}\xi }}}{\Lambda ^{4}} \)\\
 \\
\( {\hat{P}}_{2}(\xi )={\hat{P}}(\xi )\frac{e^{\frac{3\partial ^{2}}{\Lambda ^{2}\xi }}}{\Lambda ^{4}} \)\\
 \\
 \( {\hat{P}}_{3}(\xi )={\hat{P}}(\xi )\frac{e^{\frac{4\partial ^{2}}{\Lambda ^{2}\xi }}}{\xi ^{2}}\frac{\partial ^{2}}{\Lambda ^{2}}\frac{\partial ^{2}}{\Lambda ^{2}} \)\\
 \\
 ${\hat{P}}(\xi)$=${\frac{1}{1- e^{\frac{-\partial^{2}}{\Lambda^{2} \xi}}}}$ $[1-{\frac{\partial^{2}}{1-e^{\frac{-\partial^{2}}{\Lambda^{2} \xi}}}}$$\frac{e^{\frac{-\partial^{2}}{\Lambda^{2} \xi}}}{\Lambda^{2} \xi}$]
 \\
 
The terms in (3.13) involving \( {\hat{P}}_{3}(\xi ) \) can be simplified as those in (3.10)
above and adds up to a term of the same form as (3.11). The second term on the right hand side in (3.13) can
be simplified using (A.8). The residual term in (A.8) can be simplified using
(A.5) as done earlier. (3.13) can be further simplified at \( \xi  \)=1 as done in (A.10)-(A.12).\\
\\
Combining all contributions together we find \\
\\
\( \xi \frac{\partial W}{\partial \xi }|_{\xi =1,\chi =\chi =0}=<<-\frac{i}{2}\int d^{4}x\{{\mathcal{E}}_{R}^{-2}(1-\frac{\partial ^{2}}{\Lambda ^{2} })+{\hat{P}}_{3}(1)+2ig^{2}N{\hat{P}}_{2}(1){{\mathcal{M}}}\}\, \partial .A^{a}(x){\overline{\eta}} ^{a}(x) \)~\\
\( \int d^{4}y\, J^{b\mu }(y)\, \{\partial _{\mu }\eta ^{b}-gf^{bcd}{\mathcal{E}}_{A}^{2}(A_{\mu }^{c}+B_{\mu }^{c})(\eta ^{d}+\psi ^{d})\}>> \)+$\langle\langle{\it{F}}\rangle\rangle$+measure contribution.~~~~~~~~~~~~~~~~~~~~~~~(3.14)\\
\\
where the Jacobian term ${\it{F}}$ reads\\
$\it{F}$=$\frac{3 g^{2} N}{4}$[$\int$$ \frac{d^{4}k}{{(2 \pi)}^{4}} $f(${k^{2}})$$ k^{2}$] $\int $$\frac{d^{4} p}{{(2 \pi)}^{4}} $ $A_{a}$(p).$A_{a}$(-p)~~~~~~~~~~~~~~~~~~~~~~~~~~~(3.15)\\
where\\
$f(k^{2})$=[1+$\frac{k^{2} e^{\frac{k^{2}}{\Lambda^{2}}}}{\Lambda^{2}(1-e^{\frac{{k^{2}}}{\Lambda^{2}}})}$] $\frac{1}{1-e^{\frac{{k^{2}}}{\Lambda^{2}}}}$$ \frac{e^{-\frac{2 k^{2}}{\Lambda^{2}}}}{\Lambda^{4}}$~~~~~~~~~~~~~~~~~(3.16)\\
For arriving at the measure contribution, unlike other contributions, we need the form of the nonlocal BRS transformations for an arbitrary $\xi$. These are reproduced in Appendix B. We note that the nonlocal BRS and the trivial transformations do not anymore add up to a form where the following two desirable properties convenient for formulating WT identities hold:\\
(i) $\delta$$A_{\mu}$ involves the same combination that is involved in the ghost Lagrangian (ii) $\delta$$\overline{\eta}$ involves only $\partial$.A. If we define $\mu(\xi)$ with respect to either (A) nonlocal BRS of (B.1) or (B) resultant nonlocal BRS of (B.3), we have verified that the measure contribution to (3.14) cannot cancel the Jacobian contribution of (3.15).
This cancellation has to be valid in the regularized theory (i.e. for any finite $\Lambda$) and we, in particular, draw attention to the fact that 
$\langle$$\langle$$\xi$$\frac{\partial {\mu}}{\partial \xi}$$\rangle$$\rangle$ contains operators that have arbitrary order 
derivatives of A while $\it{F}$ does not.[Note the form of $\mu$ in (2.24)for $\xi$=1].\\
Finally, we wish to elaborate upon a shortcoming of the Feynman gauge treatment itself. As elaborated in 2C, the Lagrangian one starts with (of (2.19)) is actually for the case when the unrenormalized paprameter ${\xi}_{0}$=1. 
When one loop renormalization is carried out, the Lagrangian, when expressed in terms of renormalized fields, now does not retain its form of (2.19).
So when we try to extend the treatment to two loops, we have the necessity for the treatment for arbitrary $\xi$ even in this case. This, as presented here, cannot be done along the lines of this section.\    

\textbf{\textsc{}}\\
\textbf{\textsc{IV An Alternate way of regularization that preserves \( \xi  \)
independence}}\\
\\
In this section, we shall present a way of regularization that is at once
simpler and leads to WT identities that would imply the \( \xi  \) independence
of S-matrix elements (where they exist).We shall construct the relevent non-local
BRS transformation that leads to the simpler form of the WT identity. A similar regularization has already been applied to QED[7].\\
\\
We recall that the local action of the nonabelian gauge theory with an arbitrary
\( \xi  \). It is expressed as\\
\\
\( S_{\xi }=S_{F}+\Delta S \)~~~~~~~~~~~~~(4.1)\\
\\
where \( S_{F} \) is the Feynman gauge local action. \\
\\
We introduce the smeared field operators that depend only on the quadratic form
in \( S_{F} \). Hence, we have for an arbitrary \( \xi  \) \\
\\
\( {\hat{A'}}_{\mu }=({\mathcal{E}}_{F}^{-1})_{\mu \nu }A^{\nu }=e^{\frac{\partial ^{2}}{2\Lambda ^{2}}}A_{\mu } \)
~~~~~~~~~~~~~~~~~~(4.2a)\\
\\
\( {\hat{\eta}} ^{'}={\mathcal{E}}_{F}^{-1}\eta =e^{\frac{\partial ^{2}}{2\Lambda ^{2}}}\eta  \)~~~~~~~~~~~~~~~~~~~~(4.2b)\\
\\
We note \( {\mathcal{E}}_{F}={\mathcal{E}}_{\eta } \) here.\\
\\
We write down the non-local action following the same rules as those in [4]
otherwise.\\
\\
Explicitly,\\
\\
\( S^{'}_{\xi }=S^{'}_{F}+\Delta S^{'} \)~~~~~~~~~~~~~~~~(4.3)\\
\\
with\\
${S'_F}${[}A',\( \eta' ,{\overline{\eta'}}  \){]}=\( \int  \)\( d^{4}x \)\{-\( \frac{1}{2} \)\( \partial _{\mu }A'^{a}_{\nu }\, \partial ^{\mu }A'^{a\nu } \)-\( \frac{1}{2} \)\( B'^{a}_{\mu }{\mathcal{O}}^{-1}B'^{a\mu } \)-\( \partial _{\mu }{\overline{\eta'}} ^{a}\, \partial ^{\mu }{\eta'} ^{a} \)+\( {\overline{\psi'}} ^{a}{\mathcal{O}}^{-1}{\psi'} ^{a} \)\}\\
+I{[}A'+B',\( \eta' +\psi'  \),\( {\overline{\eta'}} +{\overline{\psi'}}  \){]}~~~~~~~(4.4)
\\
and\\
\\
\( \Delta S \)=\( \frac{1}{2}(1-\frac{1}{\xi })\int d^{4}x(\partial .A'^{a})^{2} \)~~~~~~~~~~(4.5)\\
\\
We note that in (4.4) the kinetic term for the auxiliary field B involves \( 
{\mathcal{O}}=\frac{{\mathcal{E}}_{F}^{2}-1}{{\mathcal{F}}_{F}} \)
that is \( \xi  \) independent. We further note that the {\bf{form}} of the relations
 between auxiliary fields (B',\( \psi' ,{\overline{\psi}}'  \)) and A',\( \eta' ,{\overline{\eta}}'  \)
is same as in Feynman gauge as $\Delta{\hat{S}}$ does not contribute to these relations.\footnote{We note that the present regularization does not preserve the properties of (2.13),(2.18). We also note that the shadow and the barred propagators in this regularization do not add up to the local one; but differ from it by "gauge terms" (that vanish as $\Lambda \longrightarrow \infty$). This does not constitute a problem however. We can look upon this as a part of existing freedom in defining the local theory itself that exists on acount of gauge invariance.} \\
\\
Now, consider the change in the effective action \( S_{\xi } \) under a field
transformation\\
\\
\( \delta {\hat{S}}_{\xi }=\delta {\mathcal{S}}[\phi ,\psi [\phi ]] \)=\( \delta {\mathcal{S}}[\phi ,\psi ]|_{\psi =\psi [\phi ]} \)~~~~~~~~~~~~~~~~~(4.6)\\
\\
Thus\\
\\
\( \delta {\hat{S}}_{\xi }=[\delta \phi \frac{\delta {\mathcal{S}}}{\delta \phi }+\delta \psi \frac{\delta {\mathcal{S}}}{\delta \psi }]_{\psi =\psi [\phi ]} \)~~~~~~~~~~~~~~~~~~~(4.7)
\\
The second term vanishes by the defining relation for \( \psi  \). \\
Thus\\
\\
\( \delta {\hat{S}}^{'}_{\xi }=\delta \phi \frac{\delta {\hat{S}}_{F}^{'}}{\delta \phi }+\delta \phi \frac{\delta \Delta {\hat{S}}^{'}}{\delta \phi } \)~~~~~~~~~~~~~~~(4.8)\\
\\
Now consider the non-local BRS transformations of the Feynman gauge non-local
action.\\
\\
\( {\hat{\delta}} A^{a}_{\mu }={\mathcal{E}}_{F}^{2}[\partial _{\mu }( \eta +\psi )^{a}-gf^{abc}(A+B)^{b}_{\mu }(\eta +\psi )^{c}]\delta \varsigma  \)~~~~~~~~~~~~~(4.9a)\\
\\
\( {\hat{\delta}} \eta ^{a}=-\frac{g}{2}f^{abc}{\mathcal{E}}_{\eta }^{2}(\eta +\psi )^{b}(\eta +\psi )^{c}\delta \varsigma  \)~~~~~~~~~~~~~~~~~~(4.9b)\\
\\
\( {\hat{\delta}} {\overline{\eta}} ^{a}=-{\mathcal{E}}_{\eta }^{2}(\partial .A^{a}+\partial .B^{a})\delta \varsigma  \)~~~~~~~~~~~~~~~~~~~(4.9c)\\
\\
We know that since \( {\hat{S}}_{F}[\phi ,\psi [\phi ]] \) is exactly of the same form
as in the Feynman gauge, it is invariant under the Feynman gauge non-local BRS
transformations of (4.9). On the other hand we find by explicit calculation\\
\\
\( \delta (\Delta {\hat{S}})=(1-\frac{1}{\xi })\int d^{4}x[{\mathcal{E}}^{2}(\partial .A^{a})]\frac{\delta {\hat{S}}_{\xi }}{\delta \eta ^{a}}\delta \varsigma  \)~~~~~~~~~~~~~~~~~(4.10)\\
\\
This change in \( \Delta {\hat{S}} \) can be canceled by an additional change\\
\\
\( {\hat{\delta'}} {{\overline{\eta}} ^{a}}=(1-\frac{1}{\xi }){\mathcal{E}}^{2}(\partial .A^{a})\delta \varsigma  \)~~~~~~~~~~~~~~~~~(4.11)\\
\\
In addition, we also note the dynamically irrelevent symmetries mentioned in
the theorem (B.4). It reads\\
\\
\( {\hat{\delta}} ^{o}A_{\mu }^{a}=[(1-{\mathcal{E}}^{2})\partial _{\mu }\eta ^{a}-{\mathcal{E}}^{2}\partial _{\mu }\psi ^{a}]\delta \varsigma  \)~~~~~~~~~~~~~~~(4.12a)\\
\\
\( {\hat{\delta}} ^{o}\eta ^{a}=0 \)~~~~~~~~~~~~~~~~~~~~~~~~~~~~~~~~(4.12b)\\
\\
\( {\hat{\delta}} ^{o}{\overline{\eta}} ^{a}=[{\mathcal{E}}^{2}(\partial .B^{a})-\frac{1}{\xi }(1-{\mathcal{E}}^{2})\partial .A^{a}]\delta \varsigma  \)~~~~~~~~~~~~~~(4.12c)\\
\\
This is an invariance of \( {\hat{S}}_{\xi } \) follows from theorem (B.4) and has
been verified by explicit evaluation.\\
\\
We add the transformations of (4.9),(4.11)and (4.12) to obtain the final non-local
BRS symmetry of the non-localized action for an arbitrary \( \xi  \). It reads,
for an arbitrary \( \xi  \):\\
\\
\( {\hat{\delta}} A_{\mu }^{a}=[\partial _{\mu }\eta ^{a}-gf^{abc}{\mathcal{E}}^{2}(A+B)^{b}_{\mu }(\eta +\psi )^{c}]\delta \varsigma  \)~~~~~~~~~~~~~~(4.13a)\\
\\
\( {\hat{\delta}} \eta ^{a}=-\frac{g}{2}f^{abc}{\mathcal{E}}^{2}(\eta +\psi )^{b}(\eta +\psi )^{c}\delta \varsigma  \)~~~~~~~~~~~~~~~~~~(4.13b)\\
\\
\( {\hat{\delta}} {\overline{\eta}} ^{a}=-\frac{1}{\xi }\partial .A^{a}\delta \varsigma  \)~~~~~~~~~~~~~(4.13c)\\
\\
This now leads to the nonlocal BRS WT identity valid for an arbitrary \( \xi  \)
of (A.4). We note here that as the Jacobian for the nonlocal trasformation (4.13) is $\xi$-independent by construction, $\mu$ can be taken to be independent of $\xi$.\\
\\
We now obtain the value of \( \xi \frac{dW}{d\xi }|_{\chi ={\overline{\chi}} =0} \), for
an arbitrary \( \xi  \), in this formulation. We note the \( \xi  \) dependence
now entirely comes from the explicit \( \xi  \) dependence of ${\Delta{{\hat{S}}}}$; since the
regulators ${\mathcal{E}}$, \( {\mathcal{O}}^{-1} \) are independent of \( \xi  \). We, thus, have\\
\\
\( \xi \frac{\partial W}{\partial \xi }=\frac{i}{2\xi }<<\int d^{4}x\, \partial .A^{a}\, {\mathcal{E}}_{F}^{-2}\, \partial .A^{a}>> \)\\
\\
The above can be effectively simplified using (A.5) {[}which now holds for an
arbitrary \( \xi  \){]}, (A.6) and (A.9) to lead to\\
\\
\( \xi \frac{\partial W}{\partial \xi }|_{J=J_{phy},\chi ={\overline{\chi}} =0}=<<-\frac{i}{2}\int d^{4}x\, [{\mathcal{E}}_{F}^{-2}\, \partial .A^{a}(x)]\eta ^{a}(x) \)~\\
\( \int d^{4}y\, J^{b\mu }(y)\{\partial _{\mu }\eta ^{b}(y)-gf^{bcd}{\mathcal{E}}^{2}(A+B)_{\mu }^{c}(\eta +\psi )^{d}\}>> \)~~~~~~~~~~~~~~~~(4.15)\\
\\
The above WT identity is the key to the $\xi$-independence of S-matrix elements (or quantities derived from them) wherever they exist. We note that in such cases, the discussion of $\xi$-independence should run entirely parallel to that in Ref.[15]. One can adopt a limiting procedure of carrying out renormalization at an off-shell point $p^2$=-${\mu}^2$ and then take the limit ${\mu}^2$$\longrightarrow$0 in the final result. 
We expect that when a similar regularization  applied in the SBGT, the resulting WT identity similar to (4.15) will lead to the $\xi$-independence of the S-matrix elements that exist.\\ 
\\
{\bf{Appendix A}}\\
\\
In this appendix, we shall derive the auxiliary equations needed in simplifying
\( \xi  \)\( \frac{\partial W}{\partial \xi } \). We consider the field transformation,
possibly nonlocal, {[}\( \epsilon  \) infinitesimal and \( \phi  \) stands
collectively for A, \( \eta  \) and \( {\overline{\eta}}  \){]}\\
\\
\( \phi \rightarrow \phi  \)+\( \epsilon  \){\it{F}}{[}\( \phi  \){]}~~~~~~~~
(A.1)\\
\\
in the field variables in W{[}J, \( \chi  \), \( {\overline{\chi}} ] \) to obtain the generalized
equation of motion:\\
\\
$\langle$$ \langle$ \( \int  \)~\( d^{4} \)x\{i\( \sum _{i} \)\( F_{i}[\phi ] \)
\( \frac{\delta {\hat{S}}_{\xi }}{\delta \phi _{i}} \)+\( \sum _{i} \)\( J_{i} \)\( F_{i} \){[}\( \phi  \){]}+${\mathcal{F}}$+\( \sum _{i} \)\( F_{i} \)\( \frac{\delta }{\delta \phi _{i}} \)ln\( \, \mu  \)\}~$\rangle$$ \rangle$=0.~~~~~~~~~~~~~(A.2)\\
\\
Here, \( \sum _{i} \)\( J_{i} \)\( F_{i} \){[}\( \phi  \){]} collectively
denotes the source terms and \( \epsilon  \)${\mathcal{F}}$ stands for the Jacobian (minus
one) for the field transformation (A.1) viz\\
 \\
${\mathcal{F}}$=\( \int  \)~\( d^{4} \)x\( \frac{\delta F[\phi ]}{\delta \phi (y)} \)\( |_{x=y} \)~~~~~~~~~~~~~~~(A.3)\\
\\
Note that the measure factor \( \mu  \)~in Feynman gauge has been chosen so
that the last two terms on the right hand side of (A.2) vanish for the (modified)
nonlocal BRS transformations of (4.13). Thus, as a special case, we have the
BRS nonlocal WT identity resulting from the surviving second term in (A.2)\\
\\
$\langle$$\langle$\( \int  \)~\( d^{4} \)x{[}\( J^{a\mu } \)\( \, {\hat{\delta}}  \)\( A_{\mu }^{a} \)+\( {{\overline{\chi}} ^{a}} \)~\( {\hat{\delta }} \)\( {{\eta} ^{a}} \)+\( {\hat{\delta}}  \)\( {\overline{\eta}} ^{a} \)~\( \chi ^{a} \){]}$\rangle$$\rangle$=0~~~~~~~~~~~~~~~~~~~~~(A.4)\\
Let P(\( \xi  \)) be any arbitrary differential operator that may depend on
\( \xi  \) but not on the fields. We operate by \( {\mathcal{E}}_{A}^{-2} \)P(\( \xi  \))\( \partial ^{\alpha } \)\( \frac{\delta }{\delta J^{p\alpha }(y)} \)\( \frac{\delta }{\delta \chi ^{c}(y)} \)
on (A.4) and put \( \chi  \)=0=\( {\overline{\chi}}  \) to obtain,\\
\\
-i\( \frac{1}{\xi } \)$\langle$$\langle$[P(\( \xi  \))\( {\mathcal{E}}_{A}^{-2} \)\( \partial . \)\( A^{p} \)\( \partial . \)\( A^{c} \)]-P(\( \xi  \))\( {\overline{\eta}} ^{c} \)\( \frac{\delta S}{\delta {\overline{\eta}} ^{p}} \)$\rangle$$\rangle$=\\
$\langle\langle \int {d}^{4}z {J}^{\mu b}(z) [{\partial}_{\mu} {\eta}^{b} - g{f}^{bcd} {{\mathcal{E}}_{A}}^{2} [{{(A+B)}_{\mu}}^{c} {(\eta+\xi)}^{d}]](z)$\\
${{\mathcal{E}}_{A}}^{-2} P(\xi) \partial. {A}^{p}(x){\overline{\eta}}^{c}(x)\rangle\rangle$~~~~~~~~~~~~~~~~~~~~~~(A.5)\\
\\
Further we note that the term of the form $\langle$$\langle$\( \int  \)\( d^{4}x \)~\( \partial ^{\mu } \)\( J_{\mu }(x) \)G{[}\( \phi  \){]}$\rangle$$\rangle$
do not contribute to the Green's function with external gauge boson lines with the physical polarization vectors since \( \epsilon  \).k=0.\\
\\
We express this by saying ~$\langle$$\langle$\( \int  \)\( d^{4}x \)~\( \partial ^{\mu } \)\( J_{\mu }(x) \)G{[}\( \phi  \){]}$\rangle$$\rangle$\( |_{phy} \)=0.~~~~~~~~~~~~~(A.6)\\
{[In using the subscript 'phy' we do not necessarily imply mass shell limit however.]}\\
We shall need a set of results derivable from (A.2)\\
\\
(I) We let F{[}\( \phi  \){]} be linear in the fields. Then ${\mathcal{F}}$ is field independent
and can be dropped while evaluating \( \xi \frac{\partial }{\partial \xi } \)
of n-point Green's functions.\\
\\
(II) With \( F^{a}_{\mu } \){[}\( \phi  \){]}=\( \partial _{\mu } \)\( {\hat{P}}_{2}(\xi ) \)\( \, \partial .A^{a} \)
(for \( \delta A_{\mu } \)) and \( F_{\eta } \)=\( F_{{\overline{\eta}} } \)=0, we then
obtain\\
\\
$\langle$$\langle$\( \int  \)\( d^{4}x \)\{i\( \partial . \)\( A^{a} \)(x)\( {\hat{P}}_{2}(\xi )\, \partial .\frac{\delta {\hat{S}}_{\xi }}{\delta A^{a}}+F_{\mu }^{a}[A]\frac{\delta }{\delta A^{a}_{\mu }} \)ln
~\( \mu  \)+i\( J^{a}_{\mu }(x)\partial ^{\mu }{\hat{P}}_{2}(\xi )\, \partial .A^{a}(x)\} \)$\rangle$$\rangle$=0~~~(A.7)\\
 \\
The measure factor at \( \xi =1 \)  has been evaluated in {[}$K_1${]} and is given
by {[}2.24{]}. Using it, we obtain, at \( \xi =1 \) and with \( J \)=\( J_{phy} \)\\
\\
$\langle\langle \int {d}^{4}x [i\partial.{A}^{a} {\hat{P}}_{2}(\xi) \partial.{\frac{\delta {\hat{S}}_{\xi}}{\delta {A}^{a}}}-{g}^{2} {f}^{acd} {f}^{bcd} {{F}^{a}}_{\mu}[A]\mathcal{M}$${A}^{b\mu}(x)]\rangle\rangle$${|_{phy}}$=0~~~~~~~~~~~~(A.8)\\
\\
(III) We let \( F_{A}=0=F_{\eta } \) and \( {F_{\overline{\eta}}}^{a}=P(\xi ){\overline{\eta}}^{a} \)
which is a linear transformation. Noting~that \( \mu  \) does not depend on
\( {\overline{\eta}}  \) and (I) above, we obtain that at \( \xi =1 \) and \( \chi ={\overline{\chi}} =0 \),\\
\\
$\langle$$\langle$\( \int d^{4}x \)P(\( \xi  \))~\( {\overline{\eta}} ^{a}\frac{\delta {\hat{S}}_{\xi }}{\delta {\overline{\eta}} ^{a}(x)} \)$\rangle$$\rangle$=0~~~~~~~~~~~~~~~~~(A.9)\\
(IV) Finally, we let \( F^{a}_{\mu } \){[}\( \phi  \){]}=\( \partial _{\mu } \)\( {\hat{P}}_{1}(\xi ) \)~\( \partial .\frac{\delta {\hat{S}}_{\xi }}{\delta A^{a}(x)} \).
For physical sources, we find\\
\\
i$\langle \langle  \int  d^{4} x  \partial . \frac{\delta {\hat{S}}_{\xi }}{\delta A^{a}} {\hat{P}}_{1}(\xi)  \partial .\frac{\delta {\hat{S}}_{\xi }}{\delta A^{a}}\rangle\rangle|_{J=J_{phy}}$=-$\langle\langle{\mathcal{F}}+\frac{\delta }{\delta A^{a}_{\mu }}ln
 \mu    F_{\mu }^{a}  \rangle \rangle$~~~~~~~~~~~(A.10)\\
\\
For \( \xi =1 \), we find that the term coming from the measure equals\\
\\
$ \int   d^{4}x   g^{2} N  \langle \langle  \partial .  {\frac{\delta {{\hat{S}}_{\xi }}}{\delta {A^{a}}}} {\hat{P}}_{1}(\xi) \mathcal{M} \partial$.${A^{a}}$$ \rangle \rangle$ +O($g^{3}$)~~~~~(A.11)\\
\\
This can be reduced further by using (A.8) to obtain\\
~\\
$\langle\langle {\frac{\delta ln\mu }{\delta A_{\mu }^{a}}}F_{\mu }^{a}\rangle\rangle |_{\xi =1,J=J_{phy}}$ =O ($g^{4}$)~~~~~(A.12) \\
\\

{\bf{Appendix B}}\\
In this appendix, we shall write down the nonlocal BRS transformations for the
case of an arbitrary \( \xi  \) following the general formalism of  [4].
We have, for the gauge fields, \\
${{\mathcal{F}}_{\mu \nu }}$= ${\eta _{\mu \nu }}$$\partial ^{2}$-(1-$\frac{1}{\xi }$)$\partial _{\mu }\partial _{\nu } \)\\
for the Euclidean formulation \( \eta _{\mu \nu } \)=diag(-1,-1,-1,-1).\\
We define\\
$({{\mathcal{E}}_{A}})_{\mu \nu }$=${(e^{\frac{\mathcal{F}}{2\Lambda ^{2}}})}_{\mu \nu }$~~~~~~~~~~~~~~~~~~~~~~~~~~~~~~~~~~~~~~~~~~~~~~~~~~~~~~~~~~(B.1)\\
We note  $\partial ^{\mu }$(${{\mathcal{E}}_{A})}_{\mu \nu }$=$e^{-\frac{\partial ^{2}}{2\Lambda ^{2}\xi }}{\partial _{\nu }}$$\equiv$${{\mathcal{E}}_{A}}^{0}$$\partial _{\nu }$.\\
For the ghost case, we continue to define  ${\mathcal{E}}_{\eta }$=$e^{-\frac{\partial ^{2}}{2\Lambda ^{2}}}$$\neq$${{\mathcal{E}}_{A}}^{0}$\\
Then the nonlocal BRS transformations read:\\
\\
\( {\hat{\delta}} A_{\mu }^{a} \) = \( {({{\mathcal{E}_{A}}^{2}})_{\mu}}^{\nu} \)\( [\partial _{\nu }(\eta ^{a}+\psi ^{a})-gf^{abc}(A^{b}_{\nu }+B^{b}_{\nu })(\eta ^{c}+\psi ^{c})]\delta \varsigma  \)\\
\\
\( {\hat{\delta}} \eta ^{a} \) = -\( \frac{g}{2} \)\( f^{abc}{{\mathcal{E}_{\eta}}}^{2}(\eta ^{b}+\psi ^{b})(\eta ^{c}+\psi ^{c})\delta \varsigma  \)~~~~~~~~~~~~~(B.2)\\
\\
\( {\hat{\delta}} {{\overline{\eta}} ^{a}}=-{{\mathcal{E}_{\eta}}}^{2}\partial _{\mu }(A^{a\mu }+B^{a\mu }) {\frac{\delta \varsigma}{\xi}}  \)\\
\\
The trivial transformations, on the other hand, read\\
\\
${{\hat{\delta}}^{o}}$${A_{\mu }}^{a}$=$\rho$$[(1-{{{\mathcal{E}}_{\eta}}^{2}})$$\partial _{\mu }\eta ^{a}$-${{\mathcal{E}}_{\eta}}^{2}$$\partial _{\mu }\psi ^{a}]$$\delta \varsigma$\\ 
\\
\( {\hat{\delta}} ^{o}\eta ^{a}=0 \)~~~~~~~~~~~~~~~~~~~~~~~~~~~~~~~~(B.3)\\
\\
${\hat{\delta}} ^{o}{\overline{\eta}} ^{a}$=$\rho$$[{{{\mathcal{E}}_{A}}^{0}}^{2}(\partial .B^{a})-(1-{{\mathcal{E}_{A}}^{0}}^{2})\partial .A^{a}]\delta \varsigma $\\
\\
where \( \rho  \) is any constant. We note that in the first of (B.2) `${\mathcal{E}}_{A}$'
appears in  $\hat{\delta} {A_{\mu}}^{a}$ , while is the first of (B.3) `${\mathcal{E}}_{\eta}$ '
appears in $\hat{\delta}^{0} {A_{\mu}}^{a}$. In the case of \( \xi  \)=1,  ${\mathcal{E}}_{\eta }$=${\mathcal{E}}_{A}$=${{\mathcal{E}}_{A}}^{0}$ .
Then with \( \rho =1, \) the first of (B.2) and (B.3) added together lead to
${\hat{\delta}}$'A  which contains the same combination of terms present in the
ghost Lagrangian and this leads to the simplification in the expression for
\( \xi \frac{\partial W}{\partial \xi } \). This no longer happens for $\xi \neq$ 1 .
Similarly, the last of (B.2) and (B.3) now contain different regulators `${\mathcal{E}_{\eta }}^{2}$ '
and `${\mathcal{E_{A}}}^{0\, 2}$' respectively. So, even with \( \rho =\frac{1}{\xi } \),
they do not lead to the cancellation of \( \partial . \)B terms. Moreover,
note that the value of \( \rho  \) needed in the first of (B.3) needed for
a `$\bf{near}$' cancellation of unwanted terms does not agree with the value of \( \rho  \)
in the last of (B.3) for a `near' cancellation of \( \partial . \)B terms.
As a result of this, for $\xi \neq$ 1, we do not have the simplified treatment
of {[}\( 4 \){]} available in the standard treatment. This holds, even
if we try to modify (B.2). \\
We could define measure \( \mu (\xi ) \) with respect to (B.2)+(B.3) with either
\( \rho =1 \) or \( \rho =\frac{1}{\xi } \) and, in either case, we find that
\( \mu (\xi ) \) must contain terms that cannot cancel the term $\it{F}$ of (3.15).

{\bf{References:}}\\
\\
1: G. 't Hooft and M. Veltman Nucl. Phys. B33, 189 (1972)\\
2: E. D. Evans et al, Phys Rev D43, 499 (1991)\\
3: See e.g, D. Z. Freedman et al, Nucl. Phys. B395, 454(1993)\\
4: G. Kleppe and R. P. Woodard, Nucl. Phys. B388, 81 (1992)\\
5: G. Kleppe and R. P. Woodard, Ann. Phys. (N.Y.) 221, 106 (1993)\\
6: See e.g. [7] and references therein\\
7: M. A. Clayton Gauge Invariance in nonlocal regularized QED. (Toronto U.) UTPT-93-14, Jul 1993. [hep-th/9307089]\\
8: S. D. Joglekar and G. Saini, Z. Phys. C.76, 343-353 (1997)\\
9: S. D. Joglekar The condition 0 $\langle$ Z $\langle$ and an intrinsic mass scale in quantum field theory. Oct1999. 9pp. and hep-th/0003077\\
10: S. D. Joglekar Understanding of the renormalization programme in a mathematically
rigorous framework and an intrinsic mass scale. Dec 1999. 18pp. and hep-th/0003104\\
11: J. Paris and W. Troost: hep-th/9607215\\
12: See e.g. E. Abers and B. W. Lee Phys. Rep. 9C(1973),1\\ 
13: S. D. Joglekar and G. Saini (Unpublished)\\
14: A. Basu and S. D. Joglekar {[}in preparation{]}\\
15: See e.g. B. W. Lee in ``Methods in Field Theory'' Les Houches 1975 Editor:
R. Balian and J. Zinn-Justin.

\end{document}